\documentclass[12pt]{amsart}
\usepackage{amsmath}

\newcommand{\cal}{\mathcal}

\newcommand{\bsab}{(\Omega_1\times\Omega_2,{\cal B}(\Omega_1\times\Omega_2))}

\newcommand{\eh}{{\cal E}\,({\cal{H})}}

\newcommand{\ran}{{\text{ran}\,}}
\newcommand{\lh}{{\cal L}({\cal H})}

\newcommand{\hi}{{\cal H}}

\newcommand{\ki}{\cal K}

\newcommand{\ip}[2]{\left\langle\,#1\,|\,#2\,\right\rangle}

\newcommand{\fii}{\varphi}

\newcommand{\sh}{\mathcal S(\hi)}
\newcommand{\tr}[1]{\mathrm{tr}\bigl[#1\bigr]}
\newtheorem{theorem}{Theorem}[section]
\newtheorem{lemma}[theorem]{Lemma}
\newtheorem{definition}[theorem]{Definition}
\newtheorem{example}[theorem]{Example}
\newtheorem{remark}[theorem]{Remark}
\newtheorem{corollary}[theorem]{Corollary}
\newtheorem{proposition}[theorem]{Proposition}

\begin{document}
\title[Coexistence]{Coexistence and Joint Measurability in Quantum Mechanics}
\author{Pekka Lahti}
\address{Pekka Lahti,
Department of Physics, University of Turku, FIN-20014 Turku, Finland}
\email{pekka.lahti@utu.fi}
\begin{abstract}
This talk is a survey of the question of joint measurability
of coexistent observables and its is based
on the monograph {\em Operational Quantum Physics} \cite{oqp}
and on the papers \cite{coex1,coex2,coex3}.

\

\noindent
A lecture 
presented at {\em Quantum Composite Systems 2002}, Ustron, Poland,
3.-7.9.2002.

\

\end{abstract}

\maketitle

\section{Introduction: coexistent observables}

The question on the possibility of measuring together two or more physical quantities 
lies at the hearth of quantum mechanics. Various notions and formulations have been
employed to investigate this issue. Von Neumann's \cite{vN32} analysis of  simultaneous measurability
of physical quantities in terms of  commutativity of the  self-adjoint operators 
representing those quantities is the starting point of much of the subsequent work.  
In particular, the investigations of Varadarajan \cite{vsv62}, Gudder \cite{Gudder68}, 
Hardegree \cite{Hardegree77}, Pulmannov\' a \cite{Pulmannova80}, and Ylinen \cite{Ylinen85}
constitute an important line of research following von Nemann's approach.

The representation of  observables as positive operator measures forces one to go beyond 
von Neumann's  framework.
Moreover, the  simultaneity of the involved measurements, that is, the fact that  the measurements
are performed at the same time point,  is, perhaps,  not the most crucial aspect of this problem.
Therefore, in that wider context, the notion of coexistence of observables has been chosen to 
describe the physical possibility of measuring together two or more quantities. 
This concept is due to G\"unther Ludwig \cite{Ludwig64} and it was further elaborated e.g. in
\cite{Ludwig67}, \cite{Hellwig}, \cite{HNeumann}, and \cite{Kraus}.
An extensive operational analysis of this notion
is presented in   \cite{Ludwig83}.

Let $\hi$ be a complex separable Hilbert space, $\lh$ the set of bounded operators on $\hi$,
$\Omega$ a nonempty set, and $\mathcal A$ a sigma algebra of subsets of $\Omega$.
We call a positive normalized operator measure $E:\mathcal A\to\lh$ an observable and
we refer to $(\Omega,\mathcal A)$ as the value space of $E$. 

Let $E,E_1$, and $E_2$ be any three observables with the value spaces $(\Omega,\mathcal A)$, $(\Omega_1,\mathcal A_1)$, 
and  $(\Omega_2,\mathcal A_2)$, and let
 $\ran(E)$ = $\{E(X)\,|\, X\in\mathcal A\}$ denote the range of $E$.

\begin{definition}\label{defcoex}
Observables $E_1:\mathcal A_1\to\lh$ and $E_2:\mathcal A_2\to\lh$ are 
coexistent  if there is an observable $E:\mathcal A\to\lh$ such that
$$
\ran (E_1)\cup\ran (E_2)\subseteq\ran (E), 
$$
that is, for each
$X\in\mathcal A_1$, and $Y\in\mathcal A_2$, $E_1(X)=E(Z_X)$, and
$E_2(Y)=E(Z_Y)$ for some $Z_X,Z_Y\in\mathcal A$.
\end{definition}

The notion of coexistence of observables is a rather
general notion and it seems to be open to characterizations only under
further specifications. They will be  studied next.

\section{Functionally coexistent observables}

\begin{definition}\label{deffcoex}
Observables $E_1:\mathcal A_1\to\lh$ and $E_2:\mathcal A_2\to\lh$ are  
functions 
of  an observable $E:\mathcal A\to\lh$,
if there are (measurable)
functions $f_1:\Omega\to\Omega_1$ and $f_2:\Omega\to\Omega_2$ such
that for each $X\in\mathcal A_1, Y\in\mathcal A_2$,
$$
E_1(X)=E(f_1^{-1}(X)),\qquad E_2(Y)=E(f_2^{-1}(Y)).
$$
In that case we say that $E_1$ and $E_2$ are functionally coexistent.
\end{definition}
As an immediate observation one has the following proposition:
\begin{proposition}
Functionally coexistent observables are coexistent.
\end{proposition}
It is an open question whether coexistent observables are functionally coexistent.
In what follows we shall investigate conditions under which coexistent observables are
functionally coexistent and we shall work out some characterisations for 
functional coexistence.  We start with another simple observation.

\begin{proposition}
Two-valued observables $E_1$ and $E_2$ are coexistent if and only if they are functionally coexistent.
\end{proposition}

\begin{proof}
To demonstrate this  fact, let $\{\omega,\omega'\}$ and
$\{\xi,\xi'\}$ be two point value sets of the observables $E_1$ and $E_2$,
with
${\rm ran}\,(E_1)= \{O,A_1,I-A_1,I\}$ and 
${\rm ran}\,(E_2)= \{O,A_2,I-A_2,I\}$,
respectively, and let $E$ be an observable
such that $E(X)=A_1$ and $E(Y)=A_2$. Consider the partition
$\mathcal R = \{X\cap Y,X'\cap Y,X\cap Y',X'\cap Y'\}$ of
the value space $\Omega$ of $E$ into disjoint $\mathcal A$-sets, and let
$1\mapsto E(X\cap Y)$,
$2\mapsto E(X'\cap Y)$,
$3\mapsto E(X\cap Y')$,
$4\mapsto E(X'\cap Y')$,
constitute a corresponding coarse-grained observable $E^{\mathcal R}$ of $E$.
The maps $f_1:1,3\mapsto \omega; 2,4\mapsto \omega'$,
and $f_2:1,2\mapsto \xi; 3,4\mapsto \xi'$,
allow one to write $A_1=E^{\mathcal R}(f_1^{-1}(\omega)) = 
E(X\cap Y)+ E(X\cap Y')$
and $A_2=E^{\mathcal R}(f_2^{-1}(\xi)) = E(X\cap Y)+E(X'\cap Y)$, showing that
the two-valued observables are functionally coexistent.
\end{proof}

Let $(\Omega_1\times\Omega_2,\mathcal A_1\times\mathcal A_2)$ denote the
product space of the measurable spaces $(\Omega_1,\mathcal A_1)$ and $(\Omega_2,\mathcal A_2)$, with
$\mathcal A_1\times\mathcal A_2=\{(X,Y)\,|\, X\in\mathcal A_1, Y\in\mathcal A_2\}$.

\begin{definition}
A positive operator  function $B:\mathcal A_1\times\mathcal A_2 \to\lh$ is a
positive operator bimeasure,  if  for each $X\in\mathcal A_1,Y\in\mathcal A_2$ 
the partial functions
\begin{eqnarray*}
&&\mathcal A_2\ni Y\mapsto  B(X,Y)\in\lh, \\
&&\mathcal A_1\ni X\mapsto B(X,Y)\in\lh,
\end{eqnarray*}
are positive operator measures.
If $B(\Omega_1,\Omega_2)=I$ we say that $B$ is a biobservable.
Observables $E_1:\mathcal A_1\to\lh$ and $E_2:\mathcal A_2\to\lh$ 
have a biobservable if there is a positive operator bimeasure 
$B:\mathcal A_1\times\mathcal A_2 \to\lh$
such that for all $X\in\mathcal A_1, Y\in\mathcal A_2$,
\begin{eqnarray*}
&&E_1(X) = B(X,\Omega_2),\\
&&E_2(Y)=B(\Omega_1,Y).
\end{eqnarray*}

\end{definition}

\

To combine observables into new observables,
biobservables or  joint observables,  to be defined below, 
some continuity properties are needed. 
It would suffice to 
assume that $\Omega$ is a Hausdorff space,  $\mathcal A = \mathcal B(\Omega)$  its Borel
$\sigma$-algebra, 
and  to require that the involde measures are Radon measures on $\mathcal B(\Omega)$
\cite{BCR84}. 
In physical applications 
the value spaces 
of observables are usually, if not always,  equipped with locally compact metrisable and
separable topologies. 
In \cite{Ludwig83} some operational justification for that structure of a value space
is also given.
The measures on the Borel $\sigma$-algebras of such spaces 
are automatically  Radon measures \cite{Halmos}.
In particular, this is  the  case for $(\Omega,\mathcal B(\Omega))$ being
the real or complex Borel spaces $(\mathbb R,{\mathcal B}(\mathbb R))$, 
$(\mathbb C,{\mathcal B}(\mathbb C))$,
or their $n$-fold Cartesian products.
To avoid the technical assumptions on Radon measures
I  assume from nowon that the value spaces of the observables are 
{\em locally compact metrisable and separable topological spaces}
and, for short,  I call them simply {\em Borel spaces}. Where this assumption is superfluous
I go on to use the notation $(\Omega,\mathcal A)$  to emphasize that no
topological assumptions are needed.

\

Let $(\Omega_1,\mathcal B(\Omega_1))$ and $(\Omega_2,\mathcal B(\Omega_2))$ 
be two Borel spaces, and
let ${\mathcal B}(\Omega_1\times\Omega_2)$ denote
the Borel $\sigma$-algebra of $\Omega_1\times\Omega_2$. 

\begin{definition}
Observables $E_1:\mathcal B(\Omega_1)\to\lh$
and $E_2:\mathcal B(\Omega_2)\to\lh$ have a joint observable
if there  is an observable
$F:\mathcal B(\Omega_1\times\Omega_2)\to\lh$ 
such that for all $X\in{\cal B}(\Omega_1), Y\in{\cal B}(\Omega_2)$, 
\begin{eqnarray}
&&E_1(X) = F(X\times\Omega_2),\\ 
&&E_2(Y)  = F(\Omega_1\times Y).
\end{eqnarray} 
\end{definition}

\begin{theorem}\label{amalgamation}
Let $(\Omega_1,\mathcal B(\Omega_1))$ and $(\Omega_2,\mathcal B(\Omega_2))$ 
be two  Borel spaces. For any two 
observables
$E_1:\mathcal B(\Omega_1)\to\lh$ and $E_2:\mathcal B(\Omega_2)\to\lh$
the following three conditions are equivalent:

\

{\rm (i)} $E_1$ and $E_2$ have a 
biobservable;

{\rm (ii)} $E_1$ and $E_2$ have a 
joint observable;

{\rm (iii)} $E_1$ and $E_2$ are functionally coexistent.

\end{theorem}

\begin{proof} 
${\rm (i)}\ \Rightarrow\ {\rm (ii)}.$ 
Let $B$ be a 
biobservable associated with $E_1$ and $E_2$. 
Then, for any $\fii\in\hi$,
the 
bimeasure $X\times Y\mapsto \ip{\fii}{B(X,Y)\fii}$ determines a unique 
measure $\mu(B,\fii)$ on 
$\bsab$
such that for all $X\in{\mathcal B}(\Omega_1)$, 
$Y\in{\mathcal B}(\Omega_2)$,
$$
\mu(B,\fii)(X\times Y) = \ip{\fii}{B(X,Y)\fii},
$$
see Theorem 1.10, p. 24, of \cite{BCR84}.
Putting $F_{\fii,\fii}(Z) =\mu(B,\fii)(Z)$ for all $\fii\in\hi, Z\in
{\mathcal B}(\Omega_1\times\Omega_2)$, one defines 
through the polarization identity and the Fr\`echet-Riesz theorem
an 
observable
$F:\mathcal B(\Omega_1\times\Omega_2)\to\lh$  
with the property 
$$
\begin{array}{ll}
&F(X\times\Omega_2) =B(X,\Omega_2) =  E_1(X),\\
&F(\Omega_1\times Y) =B(\Omega_1,Y) = E_2(Y),
$$
\end{array}
$$
for all $X\in{\mathcal B}(\Omega_1),Y\in{\mathcal B}(\Omega_2)$.\newline
${\rm (ii)}\ \Rightarrow\ {\rm (iii)}.$ 
Let now $F$ be a 
joint observable of $E_1$ and $E_2$, and let 
$\pi_1$ and $\pi_2$ be 
the coordinate projections 
$\Omega_{1} \times \Omega_{2} \rightarrow \Omega_{1}$,
$\Omega_{1} \times \Omega_{2} \rightarrow \Omega_{2}$, respectively.
Then $E_1(X) =F({\pi_1}^{-1}(X))$ and 
$E_2(Y) = F({\pi_2}^{-1}(Y))$, showing that
$E_1$ and $E_2$ are functionally coexistent.\newline
${\rm (iii)}\ \Rightarrow\ {\rm (i)}.$ 
If $E_1(X)= E({f_1}^{-1}(X))$ and  $E_2(Y)=E({f_2}^{-1}(Y))$, 
for some  observable $E:{\mathcal B}(\Omega)\to\lh$, 
and some Borel functions
$f_i:\Omega\to\Omega_i$, $i=1,2$,
then
$\ip{\fii}{B(X,Y)\fii}  := E_{\fii,\fii}(f_{1}^{-1}(X)\cap f_{2}^{-1}(Y))$,
$X \in {\mathcal B}(\Omega_1), Y \in {\mathcal B}(\Omega_2), \fii\in\hi$, 
defines a biobservable $B$ with the desired properties.
\end{proof}

\begin{example}{\rm 
Assume that the 
observables $E_1$ and $E_2$ are mutually commuting,
that is, $E_1(X)E_2(Y) = E_2(Y)E_1(X)$ for all $X\in{\mathcal B}(\Omega_1),
Y\in{\mathcal B}(\Omega_2)$.
The map  
$$
(X,Y)\mapsto E(X,Y) := E_1(X)E_2(Y),
$$
is then a biobservable. Indeed, $E(\Omega_1,\Omega_2) =I$, whereas
the positivity of $E(X,Y)$ follows from the commutativity and positivity of $E_1(X)$ and $E_2(Y)$.
The measure properties of $E_1$ and $E_2$ and the (weak) continuity
of the operator product imply that the partial functions $X\mapsto E(X,Y)$, 
$Y\in{\mathcal B}(\Omega_2)$, and $Y\mapsto E(X,Y)$, $X\in{\mathcal B}(\Omega_1)$,
are positive operator measures. Theorem~\ref{amalgamation} thus implies that any
two mutually commuting 
observables have a joint observable and
they are functionally coexistent. 
The mutual commutativity of $E_1$ and $E_2$ is, however, not necessary for any of the conditions of that theorem, as will become evident in subsequent discussion.}
\end{example}

\begin{remark}{\rm 
There is an alternative formulation of functional coexistence of obeservables,
which actually goes back to Ludwig \cite[D.3.1, page 153]{Ludwig83}
Indeed, one could say that observables $E_1$ and $E_2$ are functionally coexistence if there is an
observables $E$ and $\sigma$-homomorphisms $h_1:\mathcal B(\Omega_1)\to\mathcal B(\Omega)$
and $h_2:\mathcal B(\Omega_2)\to\mathcal B(\Omega)$ such that $E_1(X)=E(h_1(X))$ and $E_2(Y)=E(h_2(Y))$ 
for every $X$ and $Y$.  If this is the case then the
map $(X,Y)\mapsto E(h_1(X)\cap h_2(Y))$ is a bimeasure, and thus $E_1$ and $E_2$ are functionally coexistent
also in the sense of Definition \ref{deffcoex}.}
\end{remark}

\section{Regularly coexistent observables}

In a realist interpretation of quantum mechanics the notion of regular effect is an important one:
a nontrivial effect $B$ is regular if its spectrum extends both below as well as above the value 
$\frac 12$. For a further analysis of this notion the reader may consult \cite{oqp}.
Its relevance here follows from that fact that regular observables are characterized by
their Boolean range.

\begin{definition}
An observable $E:\mathcal A\to\lh$ is called regular, if for any $X\in\mathcal A$, such that $O\ne E(X)\ne I$,
$$
E(X)\not\leq \frac 12I,\ \ \ \frac 12I\not\leq E(X).
$$
\end{definition}

Clearly, an observable $E$ is regular if and only if for any  $O\ne E(X)\ne I$ neither
$E(X)\leq E(X)'$ nor $E(X)'\leq E(X)$.

\begin{lemma}\label{lemmaregular}
Let $E:\mathcal A\to\lh$ be a regular observable.
If $(A_i)_{i\in\mathbb N}\subset{\ran}\,(E)$ is a summable sequence, that is,
$A_1+\cdots +A_n\leq I$ for each $n\in\mathbb N$, 
then $\sup_{n\in\mathbb N}\{A_1+\cdots +A_n \}\in{\ran}\,(E)$.
Moreover, ${\ran}\,(E)$ is a Boolean algebra (with respect to
the order and complement inhereted from the set of effects $\eh$), and
$E$  is a Boolean $\sigma$-homomorphism $\mathcal A\to{\ran}\,(E)$.
\end{lemma}
\begin{proof}
The proof follows that of \cite[Theorem 4.1]{coex3}.
Let $A_1,A_2\in{\ran}\,(E)$ be such that $A_1+A_2\leq I$, and
assume that $A_1=E(X), A_2=E(Y)$.
Then $X=X\cap Y\cup(X\setminus (X\cap Y))$, $Y=X\cap Y\cup (Y \setminus(X\cap Y))$.
Hence $E(X\cap Y)\leq E(X)=A_1$, $E(X\cap Y)\leq E(Y)=A_2$. Since 
$A_2\leq I-A_1$, it  follows that $E(X\cap Y)\leq E(X\cap Y)'$, 
so that by the regularity assumption, $E(X\cap Y)= O$. 
Therefore $A_1=E(X_1)$ and  $A_2=E(Y_1)$,
 where $X_1:=X\setminus(X\cap Y)$,
$Y_1:= Y\setminus(X\cap Y)$ are disjoint sets. 
So we get $A_1+ A_2=
E(X_1)+E(Y_1)=E(X_1\cup Y_1)\in {\,\rm ran \,}(E)$.
We note also that $X\cap Y_1=\emptyset$ and $A_1+A_2=
E(X\cup Y_1)$. This observation will be used in the next paragraph.

Assume next that $(A_i)_{i\in \mathbb N}$ is a summable sequence in
${\rm ran \,}(E)$. Using the above argument, we find disjoint sets $X_1, X_2$
such that $A_1=E(X_1)$, $A_2=E(X_2)$. Now we proceed by induction.
Assume that we have already found disjoint sets 
$X_1,\ldots,X_{n-1}$ such that $A_i=E(X_i)$, $i=1, \ldots, n-1$. Then
$A_1+\ldots + A_{n-1}=E(X_1\cup X_2\cup  \ldots \cup X_{n-1})$.
By the summability assumption
 $(A_1+\cdots + A_{n-1})\leq A_n$.
Therefore,
there is a set $X_n\in {\cal A}$ such that $(X_1\cup \ldots \cup
X_{n-1})\cap X_n=\emptyset$, and $A_n=E(X_n)$. Thus we find a
sequence $X_i, i\in \mathbb N,$ of disjoint sets such that
$A_i=E(X_i)$, $i\in \mathbb N$. From the $\sigma$-additivity of $E$
we obtain $E(\bigcup_i X_i)=\sum_i E(X_i)=\sum_i A_i$, which shows
that ${\rm ran\,}(E)$ is closed under sums of summable sequences.

Let $E(X), E(Y)\in {\rm ran\,}(E)$. We will prove that
$$
E(X\cap Y)=E(X)\wedge_{{\rm ran\,}(E)}E(Y),
$$
 that is, 
$E:{\mathcal A}\to {\rm ran\ }(E)$ is a $\wedge$-morphism.
Evidently, $E(X\cap Y)\leq E(X), E(Y)$. Assume that for some 
$Z\in {\mathcal A}$, $E(Z)\leq E(X), E(Y)$.
We can write $Z=(Z\cap X\cap Y)\cup (Z\cap(X\cap Y)')$. Moreover,
\begin{eqnarray*} E(Z\cap(X\cap Y)') &=& E(Z\cap(X'\cup Y'))\\
&=& E((Z\cap X'\cap Y)\cup (Z\cap X'\cap Y')\cup (Z\cap X\cap Y'))\\
&=& E(Z\cap X'\cap Y)+ E((Z\cap X'\cap Y')+ E((Z\cap X\cap Y'))\\
&\leq & E(Z)\leq E(X), E(Y).
\end{eqnarray*}
But we also have 
$E(Z\cap X'\cap Y)\leq E(X')$,
$E(Z\cap X\cap Y')\leq E(Y')$, 
$E(Z\cap X'\cap Y')\leq E(X'), E(Y')$,
so that  the effects
$E(Z\cap X'\cap Y)$,  $E(Z\cap X'\cap Y')$, and $E(Z\cap X\cap Y')$
are irregular and thus equal $O$. Therefore also
$E(Z\cap(X\cap Y)')=O$.
Thus  $E(Z)=E(Z\cap X\cap Y)\leq E(X\cap Y)$. This concludes the proof
that $E(X\cap Y)=E(X)\wedge_{{\rm ran\,}(E)}E(Y)$.
By de Morgan laws one  gets the dual result:
for any $X,Y\in\mathcal A$,  
$E(X\cup Y)=E(X)\vee_{{\rm ran\,}(E)}E(Y)$.
Moreover, if the sets $X$ and $Y$ are disjoint, then
$E(X)\vee_{{\rm ran\,}(E)}E(Y)=E(X)+ E(Y)$.
Also, if $(X_i)\subset\mathcal A$ is a disjoint sequence, then
$$ 
E(\cup_{i=1}^\infty X_i) = \sum_{i=1}^\infty E(X_i)
= \bigvee_{\rm ran\,(E)}\{E(X_i)\,|\, i\in\mathbb N\}. 
$$ 

To prove that ${\rm ran\,}(E)$ is a Boolean algebra, it remains to
prove distributivity.  This follows immediately from the fact that $E$ is
a $\wedge$-morphism and  a $\vee$-morphism from a Boolean set.
\end{proof}

\begin{corollary}\label{co:bool} The range ${\rm ran\,}(E)$ of an
observable $E$ is a Boolean algebra (with the ordering inherited
from $\eh$) if and only if $E$ is regular.
\end{corollary}
\begin{proof} We have to prove the `only if' part.
Hence, assume that ${\rm ran\,}(E)$ is Boolean, and let  $E(X)$ be
an irregular element. Then $E(X)\leq E(X)'$,
which in a Boolean algebra implies that $E(X)=O$.
\end{proof}

\begin{theorem} 
For any two 
observables
$E_1:\mathcal B(\Omega_1)\to\lh$ and $E_2:\mathcal B(\Omega_2)\to\lh$,
if there is a  regular 
observable $E:\mathcal B(\Omega)\to\lh$ 
such that
$\ran (E_1)\cup\ran (E_2)\subseteq\ran (E)$,
then $E_1$ and $E_2$ are functionally coexistent.
\end{theorem}
\begin{proof}
If $E$ is regular, then from $\ran (E_1)\cup\ran (E_2)\subseteq\ran (E)$ it follows that also
$E_1$ and $E_2$ are regular. Therefore, by Lemma \ref{lemmaregular},
all the ranges $\ran (E_1),\ran (E_2),\ran (E)$ are Boolean.
From this and from the fact that $\ran (E_1)\cup\ran (E_2)\subseteq\ran (E)$
it then follows that the map
$(X,Y)\mapsto E_1(X)\land_{\ran(E)}E_2(Y)$ is a 
biobservable
of $E_1$ and $E_2$. Indeed, for a fixed $Y\in\mathcal B(\Omega_2)$,
if $(X_i)\subset\mathcal B(\Omega_1)$ is a disjoint sequence, then
\begin{eqnarray*}
E_1(\cup X_i)\land_{\ran(E)}E_2(Y)&&=\left(\sum E_1(X_i)\right)\land_{\ran(E)}E_2(Y)\\
&&=\left(\sum E(Z_{X_i})\right)\land_{\ran(E)}E_2(Y)\\
&&=\left(\bigvee_{\ran(E)}E(Z_{X_i})\right)\land_{\ran(E)}E_2(Y)\\
&&=\bigvee_{\ran(E)}\left(E(X_i)\land_{\ran(E)}E_2(Y)\right),
\end{eqnarray*}
where $(Z_{X_i})\subset\mathcal B(\Omega)$ is a disjoint sequence
such that $E(Z_{X_i})=E_1(X_i)$ (which exists since $(E(X_i))\subset\ran (E)$
is summable).
Similarly, one shows that  for  a fixed $X\in\mathcal B(\Omega_1)$, if 
$(Y_i)\subset\mathcal B(\Omega_2)$ is a disjoint sequence, then
$$
E_1(X)\land_{\ran(E)}E_2(\cup Y_i)
=\bigvee_{\ran(E)}\left(E(X)\land_{\ran(E)}E_2(Y_i)\right).
$$
Theorem \ref{amalgamation} now assures that
$E_1$ and $E_2$ are functionally coexistent.
\end{proof}

In the context of the above theorem  we say that observables $E_1$ and $E_2$ are
{\em regularly coexistent}. We may then say that regularly coexistent observables are
functionally coexistent.

\section{Projection as a value  of an observable}

Projection valued observables are known to have very special properties.
For the coexistence of two observables the fact that one of them is
projection valued implies great simplifications.
I start with quoting a well known result.

\begin{lemma}\label{commuting}
For any positive operator measure $E:\mathcal A_1\to\lh$,
if $E(X)^2=E(X)$ for some $X\in\mathcal A$, then $E(X)E(Y)=E(Y)E(X)$ for all $Y\in\mathcal A$.
\end{lemma}
\begin{proof}
Assume that $E(X)^2=E(X)$ for some $X\in\cal A$.
For any $Y\in\cal A$, $X\cap Y\subseteq Y$, so that
$E(Y) = E(Y\setminus(X\cap Y)) + E(X\cap Y)$, and
$E(X)+E(Y)-E(X\cap Y)= E(X)+E(Y\setminus(X\cap Y)) = E(X\cup Y)\leq I$.
Therefore, the effects $E(X\cap Y)$ and  $E(Y\setminus(X\cap Y))$ are below the
projections $E(X)$ and $I-E(X)$, respectively, so that
\begin{eqnarray}
E(X\cap Y) &=& E(X)E(X\cap Y)E(X),\nonumber\\
E(Y\setminus(Y\cap X)) &=& (I-E(X))E(Y\setminus(Y\cap X))(I-E(X).\nonumber
\end{eqnarray}
Therefore, $E(Y) = E(Y\setminus(X\cap Y)) + E(X\cap Y) =  E(X)E(X\cap Y)E(X)+(I-E(X))E(Y\setminus(Y\cap X))(I-E(X)$,
which gives through multiplication by $E(X)$ that $E(X)E(Y)=E(Y)E(X)$.
\end{proof}

\begin{corollary} 
Assume that 
$E_1:\mathcal A_1\to\lh$ and $E_2:\mathcal A_2\to\lh$ 
are coexistent observables. If one of them is projection valued, then they 
are mutually commuting and hence functionally coexistent.
\end{corollary}
\begin{proof}
Assume that $E_1$ is projection valued. Since $E_1$ and $E_2$ are coexistent,
Lemma~\ref{commuting} implies
that $E_1$ and $E_2$ are commuting:
$E_1(X)E_2(Y)=E_2(Y)E_1(X)$ for all $X\in\mathcal A_1,Y\in\mathcal A_2$. Thus the map
$\mathcal A_1\times\mathcal A_2\ni (X,Y)\mapsto E_1(X)E_2(Y)\in\lh$
determines a biobservable of $E_1$ and $E_2$, so that,
by Theorem~\ref{amalgamation}, 
observables $E_1$ and $E_2$ are functionally coexistent.
\end{proof}

\section{Commensurability}

For projection valued observables the following notion of commensurability,
or compatibility,
 is a further specification of the
notion of coexistence. These notions were widely used in the so-called quantum logic approaches
to quantum mechanics, see , for instance \cite{V,M,BC}.

\begin{definition} 
Projection valued observables
$E_1:\mathcal A_1\to\lh$ and $E_2:\mathcal A_2\to\lh$
are commensurable, if there is a projection valued observable $E:\mathcal A\to\lh$ such that
$\ran (E_1)\cup\ran (E_2)\subseteq\ran (E)$.
\end{definition}

\begin{theorem}
Any two projection valued  
observables
$E_1:\mathcal  B(\Omega_1)\to\lh$ and $E_2:\mathcal B(\Omega_2)\to\lh$
are coexistent if and only if they are commensurable.
\end{theorem}
\begin{proof} 
Assume that $E_1$ and $E_2$ are  coexistent projection valued observables.
By Lemma~\ref{commuting} they are mutually commuting.
Therefore,
$(X,Y)\mapsto E_1(X)E_2(Y)$
is a projection operator bimeasure, so that 
there is a joint projection valued observable $E:\mathcal B(\Omega_1\times\Omega_2)\to\lh$
 such that $E(X\times\Omega_2)=E_1(X)$ and $E(\Omega_1\times Y)=E_2(Y)$.
Thus $E_1$ and $E_2$ are commensurable.
By definition, commensurable observables are coexistent.
\end{proof}

Any two coexistent projection valued observables
$E_1$ and $E_2$ are mutually commuting:
\begin{equation}\label{commutativity}
E_1(X)E_2(Y)=E_2(Y)E_1(X)\ \ \ {\rm for\ all}\ X\in\mathcal A_1, Y\in\mathcal A_2.
\end{equation}
The pioneering result of von Neumann \cite{vN32} on commuting self-adjoint operators
gives that any two mutually commuting projection valued observables are (Borel) functions of a third
projection  valued observable. We collect these results in the following corollary.

\begin{corollary}\label{commensurability}
Let $(\Omega_1,\mathcal B(\Omega_1))$ and $(\Omega_2,\mathcal B(\Omega_2))$ 
be two  Borel spaces. For any two projection valued 
observables
$E_1:\mathcal B(\Omega_1)\to\lh$ and $E_2:\mathcal B(\Omega_2)\to\lh$
the following six conditions are equivalent:

\

{\rm(i)}\   $E_1$ and $E_2$ commute;

{\rm(ii)}\  $E_1$ and $E_2$ are commensurable;

{\rm(iii)}\  $E_1$ and $E_2$ are coexistent;

{\rm(iv)}\ $E_1$ and $E_2$ are functionally coexistent;

{\rm(v)}\  $E_1$ and $E_2$ have a 
biobservable;

{\rm(vi)}\  $E_1$ and $E_2$ have a 
joint observable.
\end{corollary}

For projection valued observables $E_1$ and $E_2$ their commutativity,  or coexistence,
or any of the above equivalent formulations, has a natural generalisation to a
partial commutativity, or partial coexistence. I shall review this question next, the basic 
results are  due to \cite{Hardegree77,Pulmannova80,Ylinen85}.

\begin{definition}
For any two projection valued observables $E_1$ and $E_2$, their commutativity domain
${\rm com}\,(E_1,E_2)$ consists of those vectors $\fii\in\hi$ for which
\begin{equation}
E_1(X)E_2(Y)\fii =E_2(Y)E_1(X)\fii
\end{equation}
for  all   $X\in\mathcal A_1,  Y\in\mathcal A_2$.
We say that $E_1$ and $E_2$ are commutative if
${\rm com}\,(E_1,E_2)=\hi$  and totally noncommutative if ${\rm com}\,(E_1,E_2)=\{0\}$.
\end{definition}

\begin{lemma}
For any two projection valued observables $E_1$ and $E_2$ their commutativity domain
${\rm com}\,(E_1,E_2)$ is a closed subspace of $\hi$ and it reduces $E_1$ and $E_2$,
that is, for any 
$X\in\mathcal A_1,  Y\in\mathcal A_2$,
\begin{eqnarray*}
&&E_1(X)({\rm com}\,(E_1,E_2))\subseteq {\rm com}\,(E_1,E_2)\\
&&E_2(Y)({\rm com}\,(E_1,E_2))\subseteq {\rm com}\,(E_1,E_2).
\end{eqnarray*}
\end{lemma}
\begin{proof}
The first claim follows since 
${\rm com}\,(E_1,E_2)$ can be expressed as the intersection
of closed subspaces,
$$
{\rm com}\,(E_1,E_2)
=\cap_{X,Y}\{\fii\in\hi\,|\, (E_1(X)E_2(Y)-E_2(Y)E_1(X))\fii = 0\}.
$$
Let $\fii\in\ {\rm com}\,(E_1,E_2)$. Then for any $Z\in\mathcal A_1$, $E_1(Z)\fii\in\ {\rm com}\,(E_1,E_2)$,
since, for all $X\in\mathcal A_1,  Y\in\mathcal A_2$,
\begin{eqnarray*}
E_2(Y)E_1(X)E_1(Z)\fii &&=E_2(Y)E_1(X\cap Z)\fii
= E_1(X\cap Z)E_2(Y)\fii\\
&& = E_1(X)E_1(Z)E_2(Y)\fii = E_1(X)E_2(Y)E_1(Z)\fii.
\end{eqnarray*}
Similarly,  one gets $E_2(Y)({\rm com}\,(E_1,E_2))\subseteq {\rm com}\,(E_1,E_2)$
for each $Y\in\mathcal A_2$.
\end{proof}

\begin{theorem}
Consider two projection valued observables $E_1$ and $E_2$ 
defined on the 
Borel spaces 
$(\Omega_1,\mathcal B(\Omega_1))$ and 
$(\Omega_2,\mathcal B(\Omega_2))$, respectively.
For any  unit vector $\fii\in\hi$,
 the following conditions are equivalent:
\

{\rm(i)}\  $\fii\in\ {\rm com}\,(E_1,E_2)$,

{\rm(ii)}\  there is a 
probability measure $\mu:\mathcal B(\Omega_1\times\Omega_2)\to[0,1]$
such that
$$
\mu(X\times Y) = \ip{\fii}{E_1(X)E_2(Y)\fii} = \ip{\fii}{E_1(X)\land E_2(Y)\fii}
$$
for all $X\in\mathcal B(\Omega_1), Y\in\mathcal B(\Omega_2)$.
\end{theorem}
\begin{proof}
The restrictions $\widetilde{E_1}$ and $\widetilde{E_2}$ of $E_1$ and $E_2$ on ${\rm com}\,(E_1,E_2)$ 
are mutually commuting spectral measures, so that, by Corollary~\ref{commensurability}, the
map $X\times Y\mapsto \widetilde{E_1}(X)\widetilde{E_2}(Y) = 
\widetilde{E_1}(X)\land\widetilde{E_2}(Y)$ 
extends to a joint projection valued observable 
$\widetilde F:\mathcal B( \Omega_1\times\Omega_2)\to\mathcal L({\rm com}\,(E_1,E_2))$.
But then, for any $\fii\in\ {\rm com}\,(E_1,E_2)$, and $X\in\mathcal B(\Omega_1), Y\in\mathcal B(\Omega_2)$,
$\widetilde F_{\fii,\fii}(X\times Y) =\ip{\fii}{\widetilde{E_1}(X)\widetilde{E_2}(Y)\fii}
= \ip{\fii}{E_1(X)E_2(Y)\fii}$, which concludes the proof.
\end{proof}

\begin{remark}{\rm 
Let $A$ and $B$ be any two self-adjoint operators in $\hi$.
According to the spectral theorem for self-adjoint operators, there are
unique spectral measures $E^A$ and $E^B$, defined on the real Borel spaces
$(\mathbb R,\mathcal B(\mathbb R))$ and taking values in $\lh$ such that $A$
and $B$ are their respective first moment operators.
By definition, $A$ and $B$ commute if and only if all their spectral projection
$E^A(X)$ and $E^B(Y)$, $X,Y\in \mathcal B(\mathbb R))$, commute.
By a well-known theorem of von Neumann \cite{vN32}, this is the case exactly when
there is a self-adjoint operator $C$ and real Borel functions $f$ and $g$ such
that $A=f(C), B=g(C)$, that is, $E^A(X)=E^C(f^{-1}(X))$ and $E^B(Y)=E^C(g^{-1}(Y))$
for all $X,Y\in \mathcal B(\mathbb R))$. We recall further that if $A$ and $B$ are
bounded self-adjoint operators, then their commutativity is equivalent with the
fact that $AB=BA$.}
\end{remark}

\section{Sequential measurements}

Let $\mathcal T(\hi)$ denote the set of trace class operators on $\hi$,
and let $\sh$ denote its subset of positive trace one operators,
the states of the quantum system associated with $\hi$.
Let $\mathcal L(\mathcal T(\hi))$ denote the set of (trace norm) bounded linear operators
on $\mathcal T(\hi)$, which is a complex Banach space with respect to the trace norm.
Let $(\Omega,\mathcal A)$ be a measurable space.
A function $\mathcal I:\mathcal A\to\mathcal L(\mathcal T(\hi))$ is an {\em instrument}
if for all $T\in\sh$ the function
$$
\mathcal A\ni X\mapsto\tr{\mathcal I(X)(T)}\in\mathbb C
$$
is a probability measure.
It follows that the function $X\mapsto E(X)$, defined through
$$
\tr{TE(X)} := \tr{\mathcal I(X)(T)},\ X\in\mathcal A, T\in\sh,
$$
is an observable $\mathcal A\to\lh$, 
the {\em associate observable} of $\mathcal I$.
It is another matter of fact that each observable $E$ is the associate
observable of some instruments $\mathcal I$; such instruments
are called $E$-compatible.

\

Consider any two instruments $\mathcal I_1:\mathcal A_1\to\mathcal L(\mathcal T(\hi))$
and $\mathcal I_2:\mathcal A_2\to\mathcal L(\mathcal T(\hi))$, and let $E_1$ and $E_2$ be their
associate observables. For each $T\in\sh$ the  function
$$
\mathcal A_1\times\mathcal A_2\ni
 (X,Y)\mapsto \mu_T(X,Y) :=
\tr{\mathcal I_1(X)\bigl(\mathcal I_2(Y)(T)\bigr)}\in [0,1]
$$
is a probability bimeasure. By the duality $\mathcal T(\hi)^*\cong\lh$, the bimeasures $\mu_T$,
$T\in\mathcal T(\hi)$, define a positive operator bimeasure
$B:\mathcal A_1\times\mathcal A_2\to\lh$ such that
$$
\tr{TB(X,Y)} = \mu_T(X,Y),
$$
for all $T\in\sh, X\in\mathcal A_1, Y\in\mathcal A_2$. 
The partial positive operator measures $E_{\Omega_1}$ and $E^{\Omega_2}$, 
associated with $\Omega_1$ and $\Omega_2$, respectively,   
are easily seen to be the observables
\begin{eqnarray}
&&E_{\Omega_1}(Y) :=  B(\Omega_1,Y) = E_2(Y),\ \ Y\in\mathcal A_2, \label{partialobservables1a}\\
&&E^{\Omega_2}(X) := B(X,\Omega_2) = \mathcal I_2(\Omega_2)^*(E_1(X)),\ \ X\in\mathcal A_1,
\label{partialobservables1b}
\end{eqnarray}
where we have used the dual transformation  
$\mathcal I_2(\Omega_2)^*:\lh\to\lh$
of the state transformation 
$\mathcal I_2(\Omega_2):\mathcal T(\hi)\to\mathcal T(\hi)$;
for all $T\in\sh, A\in\lh$
$$
\tr{T\mathcal I_2(\Omega_2)^*(A)} :=
\tr{\mathcal I_2(\Omega_2)(T)A}.
$$
We recall that using the dual transformer
$\mathcal I^*:X\mapsto\mathcal I(X)^*, X\in\mathcal  A_2$,
the associate observable $E$ of $\mathcal I$ can be expressed as
$E(X) = \mathcal I(X)^*(I)$,  $X\in\mathcal A$;  
see,  for instance  \cite{Davies}.

The above construction of biobservables shows that any two 
instruments $\mathcal I_1$ and $\mathcal I_2$,
defined on the 
Borel spaces $(\Omega_1,\mathcal B(\Omega_1))$ and 
$(\Omega_2,\mathcal B(\Omega_2))$, respectively,
give rise to a pair of 
observables for which any of
the conditions of Theorem~\ref{amalgamation} is satisfied.
These observables depend on the order in which the
instruments are applied:
\begin{eqnarray}
&&\tr{TB_{21}(X,Y)} := \tr{\mathcal I_1(X)\bigl(\mathcal I_2(Y)(T)\bigr)},
\label{partialobservables3a}\\
&&\tr{TB_{12}(X,Y)} := \tr{\mathcal I_2(Y)\bigl(\mathcal I_1(X)(T)\bigr)}.
\label{partialobservables3b}
\end{eqnarray}
In the first case these observables are those given in (\ref{partialobservables1a})
and  (\ref{partialobservables1b}), in the
second case they are given by
\begin{eqnarray}
&&
B_{12}(\Omega_1,Y) 
= \mathcal I_1(\Omega_1)^*(E_2(Y)),\ \ Y\in\mathcal B(\Omega_2),
\label{partialobservables2a}\\
&&
B_{12}(X,\Omega_2) 
 = E_1(X),\ \ X\in\mathcal B(\Omega_2).\label{partialobservables2b}
\end{eqnarray}
Usually, the sequential 
biobservables $B_{21}$ and $B_{12}$ are different.
However, it may  happen that they are the same, that is, $B_{12} = B_{21}$.
In such a case the 
observables $E_1$ and $E_2$ are, by
Theorem~\ref{amalgamation}, functionally coexistent.

\section{Joint measurability}

\

A measurement scheme for a quantum system associated with
a Hilbert space $\hi$ is a 4-tuple $\mathcal M :=\ \langle\ki,W,P,V\rangle$
consisting of a (complex separable) Hilbert space $\ki$  (describing the
measuring apparatus), a state $W\in\mathcal S(\ki)$ (the initial state of the
apparatus), an observable $P:\mathcal A\to\mathcal L(\ki)$ (the pointer observable),
and a state transformation $V:\mathcal T(\hi\otimes\ki)\to\mathcal T(\hi\otimes\ki)$
(a positive trace preserving map which models the measurement coupling).
A measurement scheme $\mathcal M$ determines an observable
$E^{\mathcal M}:\mathcal A\to\lh$ through the relation:
$$
\tr{TE^{\mathcal M}(X)} = \tr{V(T\otimes W)\, I\otimes P(X)},
\ \ T\in\sh, X\in\mathcal A.
$$
This observable is the observable measured by the scheme $\mathcal M$.
It is a basic result of the quantum theory of measurement that for each
observable $E:\mathcal A\to\lh$ there is a measurement scheme $\mathcal M$
such that $E = E^{\mathcal M}$.
A measurement scheme $\mathcal M$ also determines  a instrument
$\mathcal I^{\mathcal M}$:
$$
\mathcal I^{\mathcal M}(T) := \Pi_{\mathcal K}\bigl(V(T\otimes W)\, I\otimes P(X)\bigr),
\ \ T\in\sh, X\in\mathcal A,
$$
where $\Pi_{\mathcal K}:\mathcal T(\hi\otimes\ki)\to\mathcal T(\hi)$ is the partial trace over
the apparatus Hilbert space $\ki$.
Clearly, $E^{\mathcal M}$ is the associate observable of $\mathcal I^{\mathcal M}$.

Consider now any two  observables $E_1$ and $E_2$ (of the system with the
Hilbert space $\hi$). We say that $E_1$ and $E_2$ can be {\it measured together}
if there is a measurement scheme $\mathcal M$ and two Borel functions (pointer
functions) $f_1$ and $f_2$ such that 
\begin{eqnarray*}
&&
E_1(X) = E^{\mathcal M}(f_1^{-1}(X)), \ X\in\mathcal A_1,\\
&&
E_2(Y) = E^{\mathcal M}(f_2^{-1}(Y)), \ Y\in\mathcal A_2.
\end{eqnarray*}
It is  an immediate observation that the  observables $E_1$ and $E_2$ can be measured
together in the above sense if and only if they are functionally coexistent.

Consider next any two measurement schemes $\mathcal M_1$ and $\mathcal M_2$.
They can be applied sequentially, in either order: first $\mathcal M_1$ and
then $\mathcal M_2$, or first $\mathcal M_2$ and then $\mathcal M_1$.
This corresponds to
the application of the instruments $\mathcal I^{\mathcal M_1}$ and
$\mathcal I^{\mathcal M_2}$ one after the other, in either order.
The resulting sequential biobservables $B_{21}$ and $B_{12}$ are obtained
from Eqs (\ref{partialobservables3a}) and (\ref{partialobservables3b}).
In general, the result of such a sequential measurement depends on
the order in which the two measurements are performed.
It may happen, however,  that the measurements in question are commutative
in the sense that their sequential application is order independet.
The observables $E^{\mathcal M_1}$ and $E^{\mathcal M_2}$ determined by such
measurement schemes are functionally coexistent.
As is well known, it is highly exceptional that two measurement schemes
$\mathcal M_1$ and $\mathcal M_2$ are commutative in this sense.


\begin{thebibliography}{References}{}

\bibitem{oqp}
P. Busch, M. Grabowski, P. Lahti,
{\em Operational Quantum Physics},
{\bf LNP m31}, Springer, 2nd  corrected printing1997. 


\bibitem{coex1}
P. Lahti, S. Pulmannov\'a,
Coexistent observables and effects in quantum mechanics,
{\it Reports on Mathematical Physics} {\bf 39} (1997) 339-351.

\bibitem{coex2}
P. Lahti, S. Pulmannov\'a, K. Ylinen,
Coexistent observables and effects in convexity approach,
{\it Journal of Mathematical Physics} {\bf 39} (1998) 6364-6371.


\bibitem{coex3}
P. Lahti, S. Pulmannov\'a,
Coexistent vs. functional coexistence of quantum observables,
{\it Reports on Mathematical Physics} {\bf 47} (2001) 199-212.

\bibitem{vN32}
J. von Neumann, 
{\em Mathematical Foundations of Quantum Mechanics},
Princeton UP 1955, original German text, Springer, 1932.

\bibitem{vsv62}
V.S. Varadarajan,
Probability in physics and a theorem on simultaneous observability,
{\em Communications in  Pure and Applied Mathematics} {\bf 15}, 189-217 (1962).

\bibitem{Gudder68}
S. P. Gudder,
Joint distributions of observables,
{\em Journal of Mathematics and  Mechanics} {\bf 18}, 325-335 (1968).

\bibitem{Hardegree77}
G.M. Hardegree,
Relative compatibility in conventional quantum mechanics,
{\em Foundations of  Physics} {\bf 7}, 495-510 (1977).

\bibitem{Pulmannova80}
S. Pulmannov\' a,
Relative compatibility and joint distributions of observables,
{\em Foundations of  Physics} {\bf 10}, 641-653 (1980).

\bibitem{Ylinen85}
K. Ylinen, 
On a theorem of Gudder on joint distributions of observables,
{\em Symposium on the Foundations of Modern Physics}, pp. 691-694,
eds P. Lahti \& P. Mittelstaedt, World Scientific, 1985.

\bibitem{Ludwig64} 
G. Ludwig,
Versuch einer axiomatischen Grundlegung der Quantenmechanik und allgemeinerer physikalischer Theorien,
{\em Zeitschrift f\"ur Physik} {\bf 181} (1964) 233-260.


\bibitem{Ludwig67}
G. Ludwig,
Attempt of an Axiomatic Foundation of Quantum Mechanics and More General Thoeries, II,
{\em Communications in Mathematical Physics} {\bf 4} (1967) 331-348.

\bibitem{Hellwig}
K.-E. Hellwig,
Coexistent effects in quantum mechanics,
{\em International Journal of Theoretical Physics} {\bf 2}, 147-155 (1969).

 \bibitem{HNeumann}
H. Neumann,
Seminar notes: coexistent effects and observables,
{\em International School of Physics "Enrico Fermi"}, {\bf IL} (1970) 407-411.

\bibitem{Kraus}
K. Kraus,
{\em States, Effects, and Operations}, Springer-Verlag, Berlin 1983.


\bibitem{Ludwig83}
G. Ludwig, 
{\em Foundations of Quantum Mechanics}, Springer, 1983.



\bibitem{BCR84}
C. Berg, J. Christensen, P. Ressel,
{\em Harmonic Analysis on Semigroups}, Springer, 1984.

\bibitem{Halmos}
P. Halmos,
{\em Measure Theory}, Springer 1988, originally published in 1950.


\bibitem{V} V.S. Varadarajan,
{\it Geometry of Quantum Theory}
Springer, Berlin 1985.
First edition (in two volumes) by van Nostrand, Princeton, 1968, 1970.

\bibitem{M}
P. Mittelstaedt,
{\em Quantum Logic},
Dordrecht, Reidel, 1978.

\bibitem{BC}
E. Beltrametti, G. Cassinelli,
{\em The Logic of Quantum Mechanics},
Addison-Wesley, 1981.

\bibitem{Davies}
E.B. Davies,
{\em Quantum Theory of Open System},
Addison-Wesley, 1976,

\end{thebibliography}
\end{document}